\documentclass[sigconf, nonacm]{acmart}

\AtBeginDocument{%
  }

\setcopyright{none} 
\copyrightyear{2026}

\usepackage{amsmath,amsfonts}
\usepackage{graphicx}
\usepackage{textcomp}
\usepackage{xcolor}
\usepackage{minted}
\usepackage{algpseudocode}
\usepackage{microtype}
\usepackage{hyperref}
\usepackage{svg}
\usepackage{pgf}
\usepackage[most]{tcolorbox}
\tcbuselibrary{breakable, skins}

\newtcolorbox{boxE}[1][]{
    breakable,
    enhanced jigsaw,
    colback=gray!10, 
    colframe=black,  
    boxrule=0.5pt,   
    arc=0pt,         
    width=\linewidth,
    #1
}

\begin{document}

\title{Eidola: Modeling Multi-GPU Network Communication Traffic in Distributed AI Workloads}
\author{Ranganath Selagamsetty}
\email{selagamsetty@wisc.edu}
\orcid{0009-0007-2283-3600}
\affiliation{%
  \institution{University of Wisconsin-Madison}
  \city{Madison}
  \state{Wisconsin}
  \country{USA}
}

\author{Matthew Poremba}
\email{Matthew.Poremba@amd.com}
\orcid{0000-0001-9399-9682}
\affiliation{%
  \institution{AMD Research and Advanced Development}
  \city{Bellevue}
  \state{Washington}
  \country{USA}
}

\author{Bradford Beckmann}
\email{brad.beckmann@amd.com}
\orcid{0000-0002-5444-6521}
\affiliation{%
  \institution{AMD Research and Advanced Development}
  \city{Bellevue}
  \state{Washington}
  \country{USA}
}

\author{Joshua San Miguel}
\email{jsanmiguel@wisc.edu}
\orcid{0000-0002-6886-7183}
\affiliation{%
  \institution{University of Wisconsin-Madison}
  \city{Madison}
  \state{Wisconsin}
  \country{USA}
}

\author{Mikko Lipasti}
\email{mikko@engr.wisc.edu}
\orcid{0000-0002-8535-9244}
\affiliation{%
  \institution{University of Wisconsin-Madison}
  \city{Madison}
  \state{Wisconsin}
  \country{USA}
}




\begin{abstract}

As distributed AI workloads grow in scale, multi-GPU systems have become essential for training large models. Although techniques like kernel fusion and overlapping communication with computation help reduce delays, they also introduce irregular and transient traffic patterns that are difficult to model using existing tools. These techniques rely heavily on fine-grained synchronization and peer-to-peer communication, which place significant pressure on interconnect bandwidth and latency. 

In this work, we introduce Eidola, a scalable extension to the gem5 simulation framework that enables detailed modeling of inter-GPU communication traffic. The extension is scalable as our GPU model serves as a succinct eidolon, emulating the minimal characteristics needed for traffic modeling. Eidola uses annotated timing profiles from real applications to emulate peer-to-peer GPU writes with cycle-level precision. This allows researchers to simulate and analyze synchronization behavior across large multi-GPU configurations. The simulator supports configurable per-GPU traffic patterns and enables isolated performance analysis under different communication scenarios. 

We demonstrate Eidola’s effectiveness by reproducing variability in fused kernel execution and by implementing a SyncMon-inspired synchronization mechanism, confirming reductions in polling-related memory traffic. Our results show that Eidola provides a flexible and scalable platform for studying inter-GPU communication and supports architectural exploration in modern distributed GPU systems.

\end{abstract}


\keywords{Computer architecture research, gem5, instrumentation, multi-gpu systems}

\maketitle

\section{Introduction}

The growing demand for large-scale machine learning models has ushered in an era of multiscale graphics processing unit (GPU) computing \cite{large_scale_distributed_deep_networks_2012, squeeze_nic}, pushing system design toward exascale resources. Transformers and large language models (LLMs) continue to grow in parameter count, driving up the computational and communication demands required for training \cite{amd_at_scale_training}. While scaling out with additional GPU resources can help alleviate the training burden, diminishing returns are often observed due to kernel launch overheads, inter-GPU synchronization bottlenecks, and network-induced latencies. Simply adding more GPUs does not guarantee proportional speedups, highlighting the need for architectural and software optimizations that can better exploit available parallelism.

\begin{figure*}[htbp]
\centerline{\includegraphics[width=\linewidth]{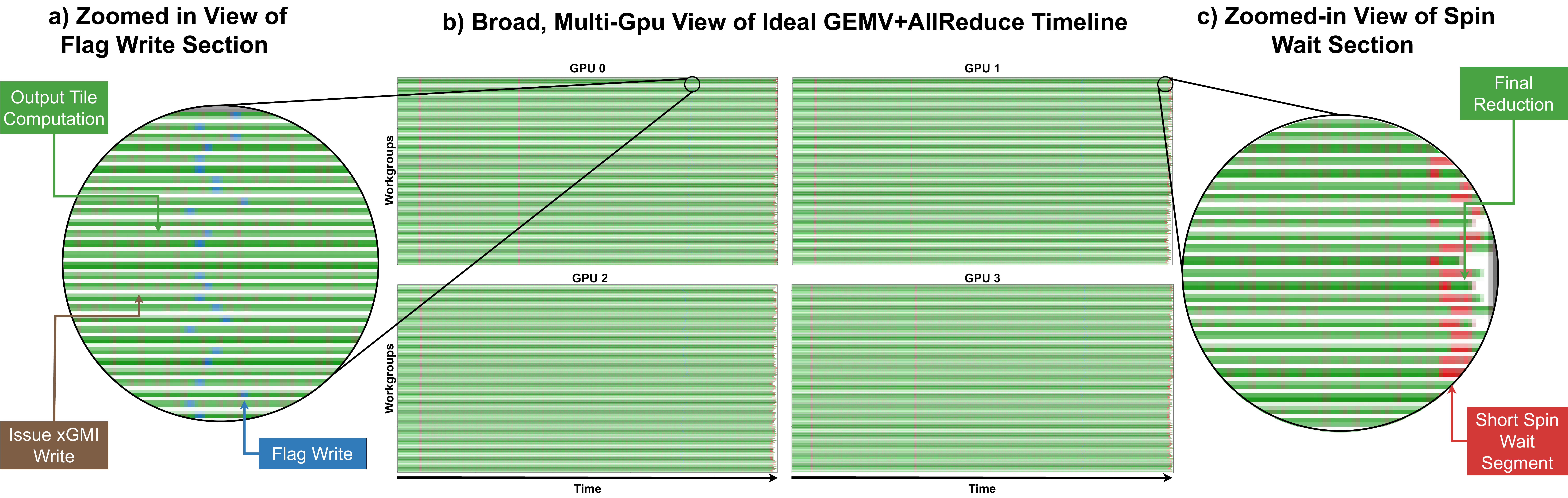}}
\caption{An ideal timing profile of the fused GEMV+AllReduce kernel on a four-GPU system. Figure \ref{fig:good_execution}\textbf{b)} shows a global view of the kernel's execution on all four GPUs. In all four GPU timing profiles, each row represents a specific workgroup's execution timeline, and time is shown on the horizontal axis. Figure \ref{fig:good_execution}\textbf{a)} highlights a period of the kernel's execution where partial output tiles needed by remote GPUs are computed, noted by the successive patterns of green segments capped by brown markers. Once all partial output tiles needed by remote GPUs are completed, a write to the a peer GPU's flag variable is issued, noted by the blue segment in Figure \ref{fig:good_execution}\textbf{a)}. The green-brown segments between the blue and red marker indicate a GPU computing partial output tiles needed by the local GPU. Finally, Figure \ref{fig:good_execution}\textbf{c)} magnifies the final portion of the kernel. The red segment indicates the time a workgroup spins on a flag variable. The final green-brown segment indicates the final output tile reduction computation. The colors and code segments are consistent with those shown in Figure \ref{fig:gemv_allreduce}.}
\Description{An ideal timing profile of the fused GEMV+AllReduce kernel on a four-GPU system. Figure \ref{fig:good_execution}\textbf{b)} shows a global view of the kernel's execution on all four GPUs. In all four GPU timing profiles, each row represents a specific workgroup's execution timeline, and time is shown on the horizontal axis. Figure \ref{fig:good_execution}\textbf{a)} highlights a period of the kernel's execution where partial output tiles needed by remote GPUs are computed, noted by the successive patterns of green segments capped by brown markers. Once all partial output tiles needed by remote GPUs are completed, a write to the a peer GPU's flag variable is issued, noted by the blue segment in Figure \ref{fig:good_execution}\textbf{a)}. The green-brown segments between the blue and red marker indicate a GPU computing partial output tiles needed by the local GPU. Finally, Figure \ref{fig:good_execution}\textbf{c)} magnifies the final portion of the kernel. The red segment indicates the time a workgroup spins on a flag variable. The final green-brown segment indicates the final output tile reduction computation. The colors and code segments are consistent with those shown in Figure \ref{fig:gemv_allreduce}.}
\label{fig:good_execution}
\end{figure*}

One promising strategy to mitigate these overheads is kernel fusion, where consecutive compute and communication operations are combined to reduce scheduling delays and improve resource utilization \cite{kernel_fusion_for_better_utilization_and_power, kernel_weaver_automatic_kernel_fusion, quantifying_benefits_of_kernel_fusion, kishore_SC24, goptx_fine_grained_kernel_fusion}. A recent example of this approach is the fused GEMV+AllReduce kernel \cite{kishore_SC24}, which merges a general matrix-vector multiplication (GEMV) with an AllReduce collective communication operation. This technique reduces launch overheads by overlapping computation with communication. Figure \ref{fig:good_execution} shows a timeline profile of the fused GEMV+AllReduce kernel's execution on a four-GPU system. In the ideal scenario, all GPUs spend the majority of the kernel execution performing arithmetic computations needed to calculate output tile results, noted by the timelines on all GPUs being predominantly green. This is ideal as no GPU spends excessive portions of their execution spin-waiting on a flag update, noted by the short red segments highlighted in Figure \ref{fig:good_execution}\textbf{c)}. 

\begin{figure*}[htbp]
\centerline{\includegraphics[width=\linewidth]{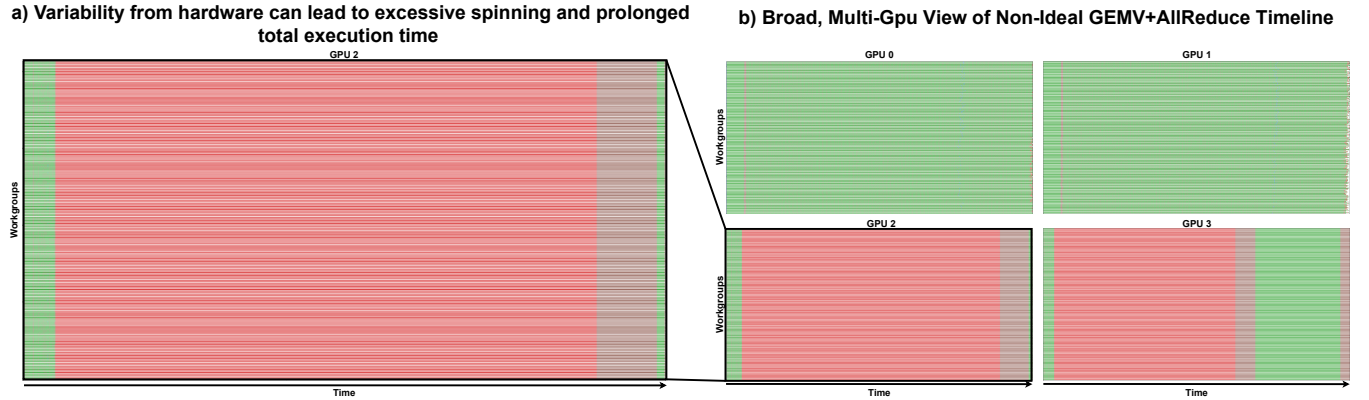}}
\caption{A non-ideal timing profile of the fused GEMV+AllReduce kernel on a four-GPU system. Like Figure \ref{fig:good_execution}\textbf{b)}, Figure \ref{fig:bad_execution}\textbf{b)} shows a global view of the kernel's execution on all four GPUs, Figure \ref{fig:bad_execution}\textbf{a)} shows a zoomed in view of GPU2's execution of the fused GEMV+AllReduce kernel. The non-ideality is apparent via the long red portions in the timeline, where all the workgroups from GPU2 spend a majority of their time spin-waiting on peer GPU flag updates.}
\Description{A non-ideal timing profile of the fused GEMV+AllReduce kernel on a four-GPU system. Like Figure \ref{fig:good_execution}\textbf{b)}, Figure \ref{fig:bad_execution}\textbf{b)} shows a global view of the kernel's execution on all four GPUs, Figure \ref{fig:bad_execution}\textbf{a)} shows a zoomed in view of GPU2's execution of the fused GEMV+AllReduce kernel. The non-ideality is apparent via the long red portions in the timeline, where all the workgroups from GPU2 spend a majority of their time spin-waiting on peer GPU flag updates.}
\label{fig:bad_execution}
\end{figure*}

However, despite workload symmetry across GPUs, variability in runtime behavior still leads to imbalanced performance due to transient network traffic and contention. Figure \ref{fig:bad_execution} shows another instance of the fused GEMV+AllReduce kernel launch to the same four-GPU system. Despite running the same kernel on the same hardware, the distribution of time spent in regions of the kernel code greatly vary across the GPUs. Like in Figure \ref{fig:good_execution}, Figure \ref{fig:bad_execution}\textbf{b)} shows the global view of the timeline profiles on all four GPUs. However, unlike Figure \ref{fig:good_execution}, GPUs two and three no longer exhibit ideal behavior. The majority of the fused GEMV+AllReduce kernel's execution time is spent spin-waiting on a flag write update from a peer GPU, noted by the large red periods in the profiles. Due to variability from runtime characteristics, the multi-GPU system experiences load imbalance across the devices, leading to prolonged kernel execution times \cite{dyn_load_balance_gpus, aeml_multi_GPU_load_balancing}. 

As fused kernels become more prevalent \cite{kernel_fusion_for_better_utilization_and_power, kishore_SC24, kernel_weaver_automatic_kernel_fusion, quantifying_benefits_of_kernel_fusion, goptx_fine_grained_kernel_fusion}, there is an urgent need for architectural tools that can accurately model and evaluate these behaviors. Existing GPU simulators, including gem5 \cite{gem5_sim} and gem5-gpu \cite{gem5_gpu}, offer cycle-level accuracy for microarchitectural research but lack native support for modeling multi-GPU systems and their associated communication dynamics. This gap limits the ability of researchers to explore GPU design trade-offs in the context of distributed training workloads.


In this work, we make the following contributions: 
\begin{itemize}
    \item A characterization exposing performance variability in fused kernel execution on mutli-GPU systems.
    \item Eidola, a scalable extension to the gem5 simulator to support multi-GPU configurations with configurable communication topologies and traffic patterns.
    \item A case study that implements the salient features of a recent work in GPU microarchitecture, SyncMon. 
    \item Scalability studies that highlight the sub-linear simulation times to evaluate multi-GPU kernels that execute on systems with tens to hundreds of GPUs. 
\end{itemize}

Our extension, \textbf{Eidola}, supports per-GPU analysis by incorporating timing and traffic profiles. In our study, these profiles were provided from real applications, but our framework can be used with synthetically generated profiles from probabilistic models. This enables researchers to evaluate the architectural impact of multi-GPU traffic on individual GPU performance in a controlled, cycle-level environment. This work provides a foundation for future architectural explorations in distributed GPU systems, especially as machine learning training pipelines continue to scale.

The rest of this paper is organized as follows. Section~\ref{sec:background} provides background on the gem5 simulator, target workloads, and the training and instrumentation cycle for ML systems. Section~\ref{sec:eidola_design} describes the design and implementation of Eidola. Section~\ref{sec:results} presents experimental results, followed by a case study in Section~\ref{sec:case_study_sync_mon}. Section~\ref{sec:related_work} discusses related work, and Section~\ref{sec:future_work} identifies directions for future research. Section~\ref{sec:conclusion} concludes with a summary of our work.

\section{Background}\label{sec:background}

This section provides the necessary background to contextualize our work. In Section \ref{sec:background_gem5}, we describe the gem5 simulation framework, which serves as the foundation for our extensions. Section \ref{sec:FusedKernelsBackgroundSection} introduces the target workload characteristics, focusing on fused computation–communication kernels commonly used in multi-GPU systems, and discuss how these kernels rely on fine-grained inter-GPU synchronization. Finally, Section \ref{sec:background_training_cycle} outlines the typical training and profiling workflow in GPU clusters, highlighting how runtime instrumentation produces the timing information leveraged by our methodology.

\subsection{The gem5 Simulator}\label{sec:background_gem5}

The gem5-gpu simulator \cite{gem5_gpu} originally extended the gem5 simulation infrastructure \cite{gem5_sim} to model heterogeneous systems that combine CPUs and GPUs. Built by coupling gem5 with GPGPU-Sim \cite{gpgpu_sim}, the tool enables full-system simulation of tightly integrated CPU-GPU architectures, supporting detailed modeling of memory hierarchies, coherence protocols, and system-level interactions. Over the past decade, gem5 has become a foundational platform for architectural research across domains including cache coherence \cite{gem5_for_cache_coherence}, memory systems \cite{pimsys_gem5}, interconnects \cite{gem5_accesys, nestar_noc_gem5}, and emerging accelerators \cite{diannao, vesper_gem5}, due to its cycle-level accuracy, modularity, and community support.

Upstream gem5 provides high-fidelity models of modern AMD GPU architectures, incorporating detailed execution pipelines, SIMT scheduling, memory subsystem behavior, and interconnect integration. These models have been rigorously validated and corroborated against empirical performance measurements from real AMD hardware, giving researchers confidence in its accuracy \cite{gem5_gpu_model_accuracy, further_gem5_gpu_modelling, modern_gpu_apps_in_gem5}. 
As such, gem5 enables researchers to evaluate architectural modifications in a realistic simulation context, making it a powerful tool for studying the performance implications of both low-level microarchitectural features and system-level interactions in heterogeneous systems. 
We build on gem5 to extend its capabilities for multi-GPU simulation and to better model the complex communication patterns observed in modern distributed machine learning workloads.

\subsection{Fused Kernels for Multi-GPU Systems}\label{sec:FusedKernelsBackgroundSection}

To mitigate the performance bottlenecks arising from kernel launch overheads and inter-GPU synchronization in distributed machine learning workloads, recent work has explored the fusion of computation and communication primitives. Notably, \cite{kishore_SC24} proposes a novel fused kernel that combines a general matrix–vector multiplication (GEMV) operation with an AllReduce collective communication step. The GEMV kernel performs a dense linear algebra operation that multiplies a matrix by a vector and accumulates the result. 
The AllReduce operation, in contrast, is a distributed communication primitive that aggregates values 
across multiple GPUs and then broadcasts the aggregated result back to all participants. 

These kernels are instrumental in auto-regressive transformer models such as ChatGPT, LLaMA2 \cite{touvron2023llama2openfoundation}, and Megatron-LM \cite{shoeybi2020megatronlmtrainingmultibillionparameter}, which rely heavily on repeated matrix operations and global synchronization of model parameters. The proposed GEMV+AllReduce fusion reduces the scheduling and synchronization delays typically encountered when these operations are launched as separate kernels. The fused kernel is especially beneficial in deep learning workloads such as transformer training, where fine-grained operations and frequent synchronizations across GPUs are common. 

\begin{figure}[!bhp]
\centerline{\includegraphics[width=\linewidth]{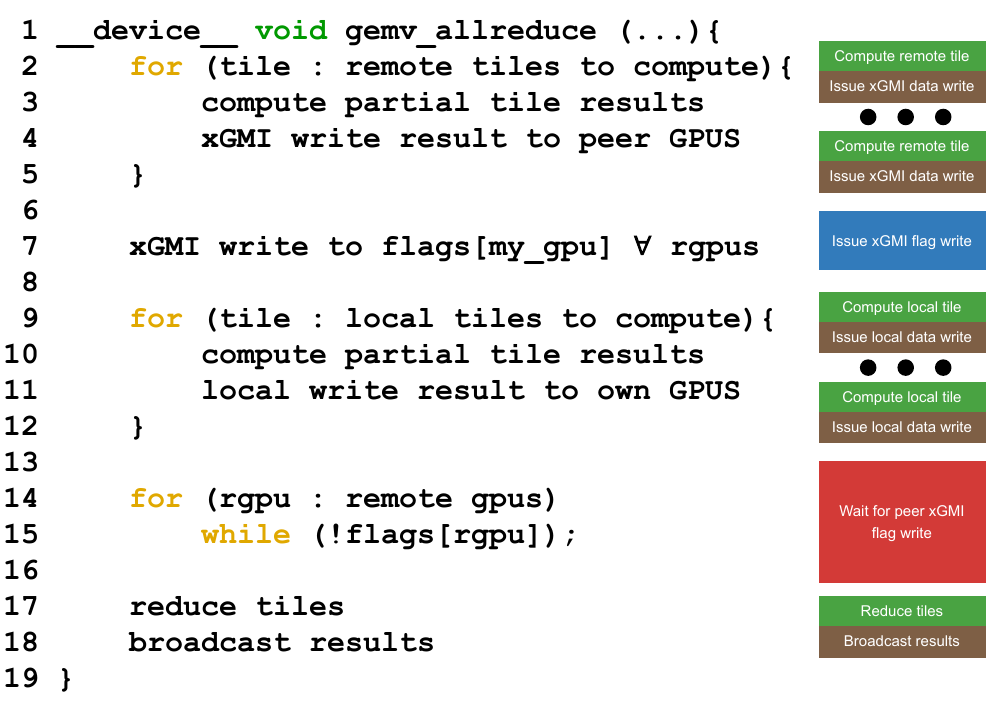}}
\caption{Pseudocode for fused GEMV+AllReduce kernel. Colored blocks on the right match color coordination from Figures \ref{fig:good_execution} and \ref{fig:bad_execution}.}
\Description{Pseudocode for fused GEMV+AllReduce kernel. Colored blocks on the right match color coordination from Figures \ref{fig:good_execution} and \ref{fig:bad_execution}.}
\label{fig:gemv_allreduce}
\end{figure}

A key architectural insight from the paper is the use of AMD’s external global memory interconnect (xGMI) to implement peer-to-peer communication between GPUs. In this fused kernel design, inter-GPU synchronization is managed through xGMI writes, where one GPU directly writes to the memory space of a peer GPU. These writes target specific flag variables, which are then polled in a lightweight control loop to gate the control flow of subsequent operations. This 
eliminates the need for costly round-trip synchronization protocols by enabling fine-grained, low-latency signaling between GPUs. As a result, GPUs can coordinate progress through the fused kernel with minimal interruption, enabling better overlap of computation and communication and reducing idle time across the cluster. The key steps in the fused GEMV+AllReduce kernel are shown in Figure \ref{fig:gemv_allreduce}. Lines \texttt{2-5} note that the kernel first computes partial sums for output tiles needed by remote GPU. Once complete, each workgroup writes a flag update to peer GPUs to indicate completion, noted by line \texttt{7}. Then, each workgroup proceeds to compute partial sums that will be reduced by the executing GPU. The results of these computation are stored locally on the executing GPU. The local GPU must now wait for all remote GPUs to compute the partial tiles needed by the local GPU, and for the flag updates to indicate so, noted by lines \ref{fig:gemv_allreduce}. Once signaled, the local GPU can perform a reduction to determine the final result for the output tile (line \texttt{17}) and broadcasts the results to all GPUs (line \texttt{18}).

Inter-GPU communication through xGMI operates within the GPU memory hierarchy, which is organized into several caching layers and implements directory-based cache coherence. Each compute unit (CU) contains private L1 caches that service local memory accesses and instructions. These L1 caches are connected to a shared L2 cache, which may be partially or fully shared among multiple CUs depending on the GPU architecture. The L2 cache maintains coherence among the L1 caches and interfaces with the directory, which serves as the global coherence manager for the device. The directory tracks ownership and sharing states for cache lines, coordinating memory requests between the GPU’s internal caches, device memory, and external agents such as peer GPUs or CPUs.

In \cite{kishore_SC24}, the synchronization elements, or flag variables used for inter-GPU coordination, are designated as non-cacheable memory locations. Treating these regions as non-cacheable ensures that updates are observed immediately across devices without requiring explicit invalidation or coherence traffic through the L1 or L2 caches. When a peer GPU issues an xGMI write to a polling GPU’s memory space, the request is directed through the xGMI fabric to the target GPU’s cache directory. The directory records the update atomically with respect to any pending polling reads from the local GPU, ensuring that subsequent accesses observe the most recent flag value.

This behavior is facilitated by rocSHMEM, a GPU-resident communication runtime developed by AMD and AMD Research to provide GPU-centric networking through an OpenSHMEM-like interface \cite{rocshmem_github}. rocSHMEM exposes one-sided communication primitives such as \texttt{put}, \texttt{get}, and \texttt{atomic} operations directly from GPU kernels, allowing threads to perform fine-grained data movement and synchronization without host intervention. It implements these primitives atop the xGMI interconnect, enabling intra-kernel communication and synchronization across GPUs at device speed. The library allocates a single symmetric heap across all participating GPUs, residing in each device’s memory space, which ensures a uniform address layout and simplifies pointer arithmetic in distributed kernels. By integrating communication into the GPU execution model, rocSHMEM reduces code complexity and allows for tighter coupling of computation and communication phases, where progress signaling and data exchange occur continuously within a single launch.

This mechanism provides an efficient and coherent signaling pathway between GPUs. By avoiding round-trip synchronization through the CPU or host memory subsystem, xGMI writes enable fine-grained coordination at device speed. The write completes transparently at the cache directory level, simplifying coherence management in the polling GPU’s memory hierarchy while maintaining the atomicity and ordering guarantees required for synchronization. As a result, GPUs can exchange progress signals through low-latency peer-to-peer writes, supporting tightly coupled fused kernels that overlap computation and communication effectively.

While this mechanism provides programmer-transparent communication between multiple GPUs, it also introduces new architectural complexity that is difficult to analyze using existing tools. The precise timing of xGMI transactions, cache directory updates, and polling behaviors directly influence GPU utilization, synchronization efficiency, and the ability to hide communication latency. Small variations in these interactions can lead to measurable differences in overall throughput and scaling behavior, especially in tightly coupled workloads such as fused GEMV+AllReduce \cite{kishore_SC24}. Capturing these nuances requires simulation frameworks that model both intra-GPU microarchitectural events and inter-GPU communication traffic at fine granularity.

This tightly integrated approach to fused computation and synchronization highlights the importance of modeling both microarchitectural behavior and interconnect traffic in modern multi-GPU systems. However, cycle-level simulators have yet to fully support such fused communication patterns, motivating our extension of gem5 to capture these interactions in a realistic and configurable simulation environment.

\subsection{Training Cycle on GPU Clusters}\label{sec:background_training_cycle}

Training modern machine learning and artificial intelligence models at scale follows an iterative life-cycle that combines model design, large-scale deployment, and continuous performance tuning. A typical workflow begins with model construction and data preparation, followed by distributed training across multi-GPU or multi-node clusters. Each training iteration executes a forward pass to compute activations and a backward pass to propagate gradients, both of which are highly parallelized across many GPUs. Distributed data-parallel training frameworks such as PyTorch Distributed \cite{pytorch_distributed}, Horovod \cite{horovod}, and DeepSpeed \cite{deepspeed_microsoft} manage the synchronization of gradients through collective communication operations, which aggregate partial results across devices.

\begin{figure}[htbp]
\centerline{\includegraphics[width=\linewidth]{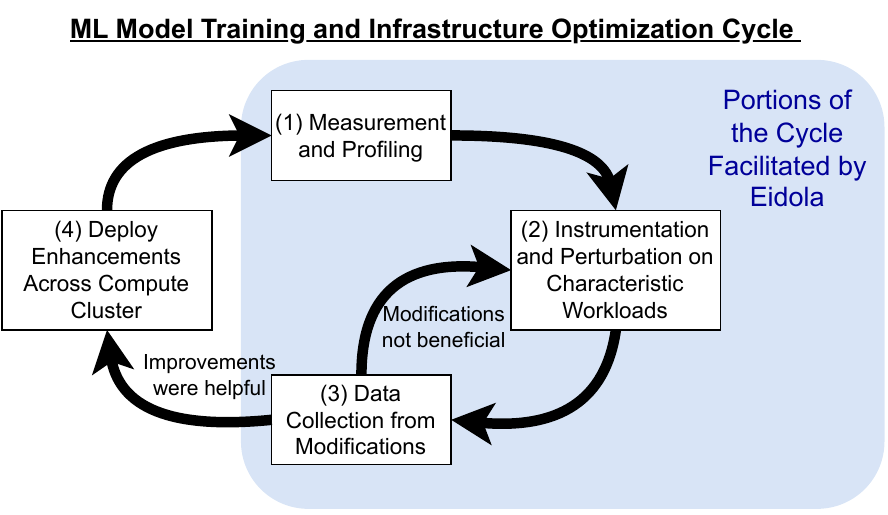}}
\caption{Iterative optimization cycle for large-scale GPU cluster training. The workflow consists of (1) measurement and profiling of baseline workloads, (2) instrumentation and controlled perturbation of representative kernels, (3) data collection and analysis of resulting performance characteristics, and (4) deployment of validated optimizations across the cluster. The stages highlighted in blue correspond to phases where Eidola facilitates analysis by enabling controlled replay and modeling of inter-GPU communication behavior.}
\Description{Iterative optimization cycle for large-scale GPU cluster training. The workflow consists of (1) measurement and profiling of baseline workloads, (2) instrumentation and controlled perturbation of representative kernels, (3) data collection and analysis of resulting performance characteristics, and (4) deployment of validated optimizations across the cluster. The stages highlighted in blue correspond to phases where Eidola facilitates analysis by enabling controlled replay and modeling of inter-GPU communication behavior.}
\label{fig:ml_training_cycle}
\end{figure}

Large-scale systems such as NVIDIA DGX SuperPODs \cite{nvidia_dgx_superpod} or AMD Instinct\texttrademark~MI300 clusters \cite{amd_mi300_clusters} rely on high-bandwidth, low-latency interconnects to mitigate the inter-GPU communication as model sizes and datasets complexities grow. However, scaling inefficiencies persist due to factors such as kernel launch latency, load imbalance across devices, and variable communication delays introduced by the underlying network topology \cite{aeml_multi_GPU_load_balancing}. Even in seemingly homogeneous environments, contention within the interconnect fabric and subtle runtime differences among GPUs can lead to performance variability across nodes \cite{data_movt_amd_mi250_cluster, vortex_vldb_2025}. These sources of non-determinism affect convergence time, reduce throughput, and complicate scheduling and performance prediction.

Profiling is an essential step in addressing these challenges \cite{google_wide_profiling, digital_continuous_profiling}. During distributed training, runtime profiling tools such as NVIDIA Nsight\texttrademark~Systems \cite{nvidia_nsight_systems}, ROCm Profiler \cite{rocm_profiler_2022}, and PyTorch Profiler \cite{pytorch_profiler_2021} are used to measure GPU utilization, communication latency, and kernel execution timing \cite{picongpu_using_nsight_systems, cpu_gpu_profiling_rocm_profiler, prof_generation_using_rocprof, prof_monitor_dl_training_pytorch_profiler}. These measurements help developers identify inefficiencies such as unbalanced workloads, excessive synchronization waits, or communication bottlenecks. The collected data also inform strategies for hyperparameter tuning, data partitioning, and kernel fusion, which can significantly improve hardware utilization and reduce total training time. Profiling further provides the empirical foundation for simulation-based performance modeling, allowing researchers to explore new architectural optimizations before hardware implementation. This can be in the form of tuning a synthetic traffic generator during simulation \cite{synfull_network_traffic_generator, gmap_gpu_mem_traffic_sim, metoo_stoch_mem_traffic_generator, stoch_synth_trace_generation}, or, as done in this work, replaying real-life traffic patterns \cite{mocktails_replay_mem_behavior, cinda_instr_data_mem_traffic_cloning}.

Figure~\ref{fig:ml_training_cycle} illustrates this iterative optimization cycle. After an initial training deployment, developers first perform measurement and profiling to identify performance bottlenecks at both the kernel and system levels. These observations motivate targeted instrumentation of characteristic workloads, where specific communication or synchronization behaviors are isolated and, in some cases, deliberately perturbed to study their impact. The resulting execution traces are then collected and analyzed to understand how low-level effects—such as inter-GPU communication latency or synchronization delay—propagate to overall training performance. Once validated, optimizations derived from this process are deployed across the full compute cluster, completing the cycle.

Eidola is designed to facilitate the first three stages of this workflow. By leveraging lightweight timing profiles and enabling controlled replay of inter-GPU communication events, Eidola allows researchers to systematically study the impact of communication timing and synchronization behavior without requiring repeated execution on large-scale hardware. In particular, the ability to perturb communication timing in a controlled, cycle-level simulation environment enables fine-grained analysis that complements traditional profiling tools. This makes Eidola a practical bridge between empirical measurement and architectural exploration, supporting rapid iteration on communication-centric optimizations before deployment at scale.

By accurately capturing the timing and synchronization patterns observed in real multi-GPU workloads, profiling enables researchers to construct realistic models of distributed training behavior. These models are critical for evaluating architectural proposals that target inter-GPU communication, memory hierarchy design, and collective operation efficiency. As GPU clusters continue to scale toward exascale performance, profiling and modeling remain central to understanding and optimizing the interaction between computation and communication at every layer of the AI training stack.

\section{Eidola}\label{sec:eidola_design}

This section describes the design and implementation of Eidola within the gem5 simulation framework. We begin in Section~\ref{SetupKernelSection} by introducing the use of a setup kernel, which is launched prior to the kernel under study to prime simulator-side data structures with timing and communication information. We then discuss additional design considerations in Section~\ref{sec:other_design_options}, outlining alternative approaches we explored during development and highlighting key limitations that future users of gem5 should be aware of. Together, these sections detail both the core mechanisms of Eidola and the practical insights gained during its integration.

\subsection{Registering Writes via Setup Kernel}\label{SetupKernelSection}

In the typical training life-cycle of large-scale GPU workloads, runtime profiling is an integral part of performance tuning and optimization. The timing profiles gathered during this process, or synthetically generated patterns, can be directly leveraged to inform simulation studies. Specifically, these profiles capture inter-GPU communication behavior, including peer-to-peer writes that occur during fused computation–communication kernels. By annotating each write operation with its corresponding timestamp, the collected traces provide detailed insight into when and how GPUs exchange data.

For the fused GEMV+AllReduce kernel examined in this work, only the timestamps of peer-to-peer write operations are required to reconstruct realistic inter-GPU synchronization behavior. For other distributed workloads, users can choose to collect timing information with greater granularity, including additional communication events or memory operations, depending on the desired level of modeling fidelity. This flexibility allows the simulator to accommodate a broad range of use cases, from fast, approximate modeling to highly detailed, cycle-level analysis of multi-GPU communication dynamics.

\begin{figure}[htbp]
\centerline{\includegraphics[width=\linewidth]{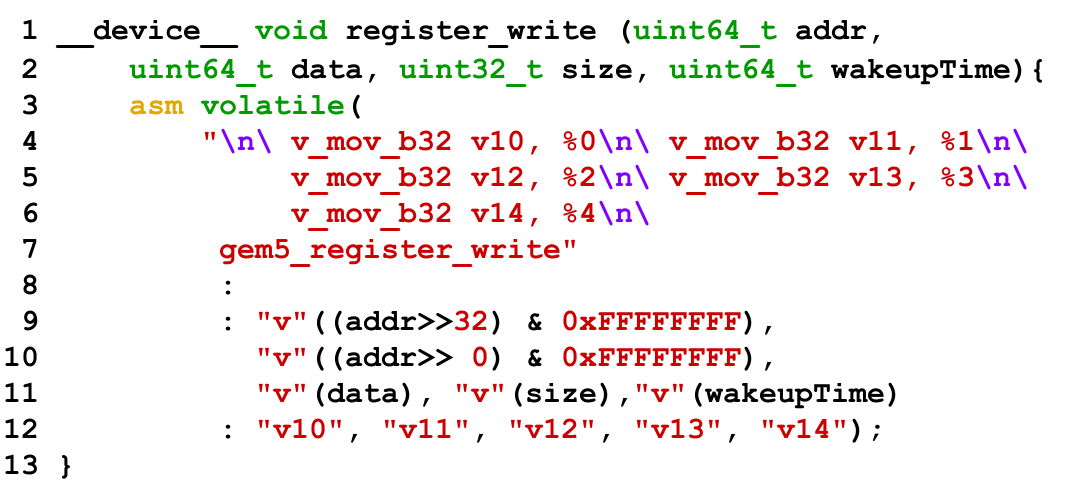}}
\caption{Code snippet shown implementation details of proposed setup kernel.}
\Description{Code snippet shown implementation details of proposed setup kernel.}
\label{fig:register_write}
\end{figure}

To inject these communication events into the simulation, we introduce the new GPU pseudo op, \texttt{register\_write}, shown in Figure \ref{fig:register_write}. This function was inserted into the application before the launch of the main computation kernel and recognized by the simulator as part of a setup phase. During simulation, this setup kernel executes in functional mode, not in detailed timing mode. The sole purpose of the setup kernel (or kernels if multiple writes to register) is to preload the simulator with information about upcoming inter-GPU writes that will be enacted during detailed simulation of the main kernel.

The \texttt{register\_write} function takes four parameters: \texttt{addr}, \texttt{data}, \texttt{size}, and \texttt{wakeupTime}. The \texttt{addr} parameter specifies the destination address for the emulated write, \texttt{data} holds the value to be written, and \texttt{size} indicates the width of the write in bytes (ranging from one to eight bytes). The \texttt{wakeupTime} parameter defines the time offset, in nanoseconds, after kernel launch at which the write should be issued. At runtime, these parameters are read by the simulator and stored in a queue, sorted by \texttt{wakeupTime}. These timestamps are converted into cycles based on the device clock frequency defined in the gem5 configuration.

We extend the simulator with a data structure to maintain all pending writes in a write tracking table (WTT), implemented as a priority queue sorted by \texttt{wakeupTime}. When a write is registered, it is inserted into the WTT at a position determined by its \texttt{wakeupTime}, ensuring that the earliest writes are processed first. During simulation of the main kernel, the head of the WTT is polled on every simulated cycle tick. If the current simulated time is less than the \texttt{wakeupTime} of the head entry, the simulator simply advances to the next cycle, incurring negligible runtime overhead (only a single comparison per cycle). When the simulated time meets or exceeds the \texttt{wakeupTime}, all write entries at the head of the queue with the same timestamp are popped and enacted as xGMI writes. These events are processed in constant time, as the number of simultaneous writes is typically small and bounded by the number of participating GPUs. This priority queue-based design ensures both temporal accuracy and efficiency, allowing users to register writes in arbitrary order. The means sequential calls to \texttt{register\_write()} need not correspond to the chronological order of their execution. By decoupling registration from enactment, this mechanism allows flexible modeling of inter-GPU communication patterns with minimal simulation overhead.

As mentioned above, when the main kernel enters detailed timing mode, the simulator begins polling the WTT on every simulated cycle. Once the current simulation time reaches the wakeupTime associated with the next pending event, the simulator issues the corresponding xGMI write on behalf of the emulated GPU. The write transaction completes at the cache directory level, accurately reflecting the hardware behavior described in Section \ref{sec:FusedKernelsBackgroundSection}, where peer-to-peer writes update non-cacheable memory regions atomically and without round-trip acknowledgment. Upon completion, the memory state of the receiving GPU is updated to reflect the new flag value. When the receiving GPU subsequently polls this address the updated data is detected and propagated to the requesting CU.

This implementation provides a faithful simulation of the low-latency, one-sided communication semantics employed in rocSHMEM \cite{rocshmem_github}, where intra-kernel synchronization primitives rely on direct xGMI writes rather than host-mediated messaging. By modeling these events at cycle granularity and maintaining an efficient WTT structure, the simulator achieves both temporal fidelity and scalability, enabling the study of tightly coupled multi-GPU workloads that overlap computation and communication within a single fused kernel execution. The flexibility of Eidola's write scheduling methodology allows this framework to be useful for a variety of memory traffic generation schemes, including synthetic, \cite{synfull_network_traffic_generator, gmap_gpu_mem_traffic_sim, metoo_stoch_mem_traffic_generator, stoch_synth_trace_generation}, as well as replicatory \cite{mocktails_replay_mem_behavior, cinda_instr_data_mem_traffic_cloning}.

\subsection{Other Design Considerations} \label{sec:other_design_options}

\subsubsection{CPU Orchestration vs Setup Kernel}

Before adopting the GPU pseudo op approach, we initially explored using CPU pseudo ops to emulate GPU writes. This method involved instrumenting the application with CPU threads that issued pseudo ops during the execution of the main GPU kernel. It was appealing due to its simplicity, minimal simulator modifications, and consistency with gem5's original heterogeneous CPU-GPU simulation model.

However, this approach proved impractical. In gem5, CPU and GPU components are serialized, not concurrent as in real hardware. Enabling CPU threads to issue writes while the GPU kernel runs would require frequent simulator context switches. These switches are configured based on host machine parameters such as core count, clock frequency, and microarchitecture, which are not part of the simulated system. This made the timing behavior unreliable and hard to reproduce for cycle-level studies. We advise against using CPU pseudo ops for emulating GPU behavior in gem5. While simpler, this method introduces nondeterminism in multi-GPU simulation contexts\footnote{gem5’s GPU model typically executes on a KVM-backed CPU, the CPU-side portions of the application simulation depend on the host machine’s hardware characteristics (e.g. core count, clock behavior, scheduling policy, etc.). As a result, GPU–CPU interaction can exhibit nondeterministic timing and context-switch behavior that does not reflect the modeled system.}. Prior work examined gem5 simulation scalability in multi-CPU systems \cite{cpu_scaling_gem5}, but we avoided this implementation as naively partitioning GPUs between functional and timing models did not improve simulation time due to their larger, high-fidelity design.

\subsubsection{Event Queues vs Write Tracking Table}

Earlier in Section \ref{SetupKernelSection}, we described our modifications to the gem5 simulator that introduced a new GPU pseudo-op to interact with the custom WTT data structure. The WTT provides an intuitive interface for scheduling and emulating timed inter-GPU writes, implemented as a priority queue whose head is polled each simulation cycle. While this design offers transparency and ease of debugging (all pending and enacted writes can be directly inspected during runtime), it does incur a polling overhead due to the per-cycle check on the queue head. This cost is minute, since the poll reduces to a constant-time comparison between the current simulation tick and the wakeup time of the head entry. In the common case, the scheduled write lies in the future, so the simulator performs only this O(1) check without additional processing.

In the native gem5 simulation framework, timing-sensitive events such as memory transactions, DMA operations, and synchronization events are typically scheduled using event queues. These queues allow the simulator to efficiently manage future events without active polling, as each event is inserted with an associated timestamp and automatically triggered when its scheduled time arrives. Leveraging this existing event-driven infrastructure would eliminate the need for explicit polling of the WTT, thereby improving runtime efficiency and scalability for large-scale multi-GPU simulations.

\begin{figure*}[!htbp]
\centerline{\includegraphics[width=\linewidth]{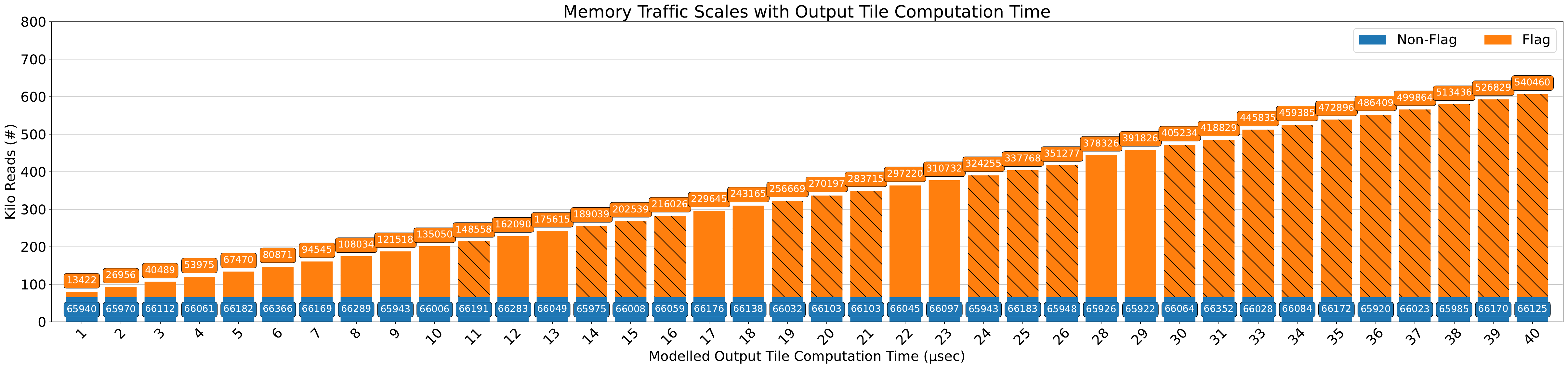}}
\caption{Fine-grained control of simulated multi-GPU communication traffic. Horizontal axis shows sweep of \texttt{wakeupTime} parameter for registered writes in setup kernel. Vertical axis shows number of read requests issued by the fused GEMV+AllReduce kernel application. Non-flag related read requests are shown in blue (patterned with forward slashes), while those issued from the spin-wait are shown in orange (patterned with back slashes).}
\Description{Fine-grained control of simulated multi-GPU communication traffic. Horizontal axis shows sweep of \texttt{wakeupTime} parameter for registered writes in setup kernel. Vertical axis shows number of read requests issued by the fused GEMV+AllReduce kernel application. Non-flag related read requests are shown in blue (patterned with forward slashes), while those issued from the spin-wait are shown in orange (patterned with back slashes).}
\label{fig:configurability}
\end{figure*}

Integrating the WTT functionality into gem5’s event queue subsystem represents a promising direction for future work. This approach would maintain the same temporal fidelity while reducing the simulator’s runtime overhead, particularly in workloads that register thousands of inter-GPU writes per kernel. Moreover, using the native event queues would align the GPU pseudo-op mechanism with gem5’s established scheduling paradigm, simplifying maintenance and improving interoperability with other timing events in the simulation pipeline. While the current WTT-based implementation prioritizes debuggability and transparency of simulator state, future releases may incorporate an event-driven backend to further enhance performance and reproducibility.

\section{Results}\label{sec:results}

Our simulation framework enables full instrumentation of inter-GPU network traffic patterns through annotated, timestamped write instructions. This design provides researchers with precise control over when communication events occur relative to GPU kernel execution. To demonstrate this capability, we evaluate how varying the \texttt{wakeupTime} parameter of our registered write impacts observed memory traffic in simulation.

Figure \ref{fig:configurability} illustrates this behavior by sweeping the wake-up time of the emulated write from \texttt{0} to \texttt{40} microseconds. A wake-up time of \texttt{0} indicates the emulated write is issued at the start of the main kernel execution, while higher values delay the emulated write correspondingly. The vertical axis shows the total number of memory read requests observed during simulation. Each bar is divided into two categories: non-flag reads (shown in blue), which represent general memory traffic, and flag reads (shown in red), which result from spin-wait loops polling for the flag update. These polling reads correspond to the flag-waiting behavior on lines \texttt{14-15} of the fused kernel code shown in Figure \ref{fig:gemv_allreduce}.

\begin{table}[hbp]
  \centering
  \caption{Simulator and Application Configurations}
  \label{table:sim_app_configs}
  \begin{tabular}{|l|c|}
    \hline
    \multicolumn{2}{|c|}{Simulation Configurations} \\
    \hline
    Number of CUs in simulated GPUs & 4\\
    Number of emulated GPUs & 3\\
    Workgroups launched per GPU & 208\\
    \hline
    \hline
    \multicolumn{2}{|c|}{Application Configurations} \\
    \hline
    Input Matrix dimension M & 256 \\
    Input Matrix dimension K & 8192\\
    Input Matrix dimension N & 1\\
    \hline
\end{tabular}
\end{table}

As expected, the number of flag reads increases linearly with the delay introduced by \texttt{wakeupTime}. This confirms that the simulator faithfully reproduces fine-grained timing behavior consistent with GPU spin-wait synchronization. The clear separation between non-flag and flag traffic also highlights the utility of our method in isolating and analyzing communication-driven memory access patterns.

\begin{figure*}[htbp]
\centerline{\includegraphics[width=\linewidth]{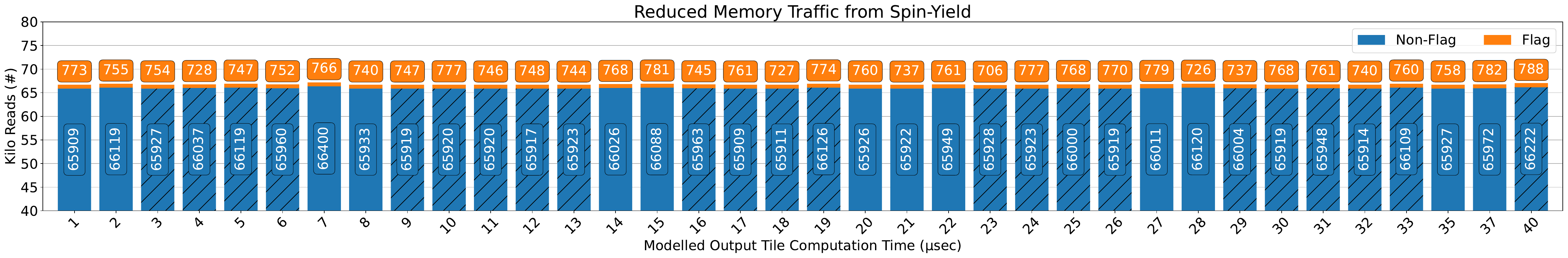}}
\caption{Read request reduction when adopting a spin-yield pattern synchronization pattern. Axes and color format is consistent with Figure \ref{fig:configurability}.}
\Description{Read request reduction when adopting a spin-yield pattern synchronization pattern. Axes and color format is consistent with Figure \ref{fig:configurability}.}
\label{fig:syncmon_results}
\end{figure*}

This experiment validates that our instrumentation and simulator extensions provide a transparent and flexible way to explore the performance impact of inter-GPU communication in distributed workloads. Table \ref{table:sim_app_configs} summarizes the simulator configuration and application parameters used in Figures \ref{fig:configurability} and \ref{fig:syncmon_results}. 

\section{Case Study: SyncMon}\label{sec:case_study_sync_mon}

To demonstrate the utility and flexibility of our simulation methodology, we implement the key features of a recent GPU microarchitecture \cite{ifp_sync_mon}. This work introduced the synchronization monitor (SyncMon), which enables fine-grained synchronization support in GPU hardware. The SyncMon mechanism, along with associated hardware logic, allows GPU threads to use monitor/wait semantics similar to the \texttt{monitor}/\texttt{mwait} instruction pair in CPUs. Instead of continuously polling in a tight spin-wait loop, threads can now yield GPU compute resources while waiting, adopting a spin-yield synchronization pattern.

\begin{figure}[htbp]
\centerline{\includegraphics[width=\linewidth]{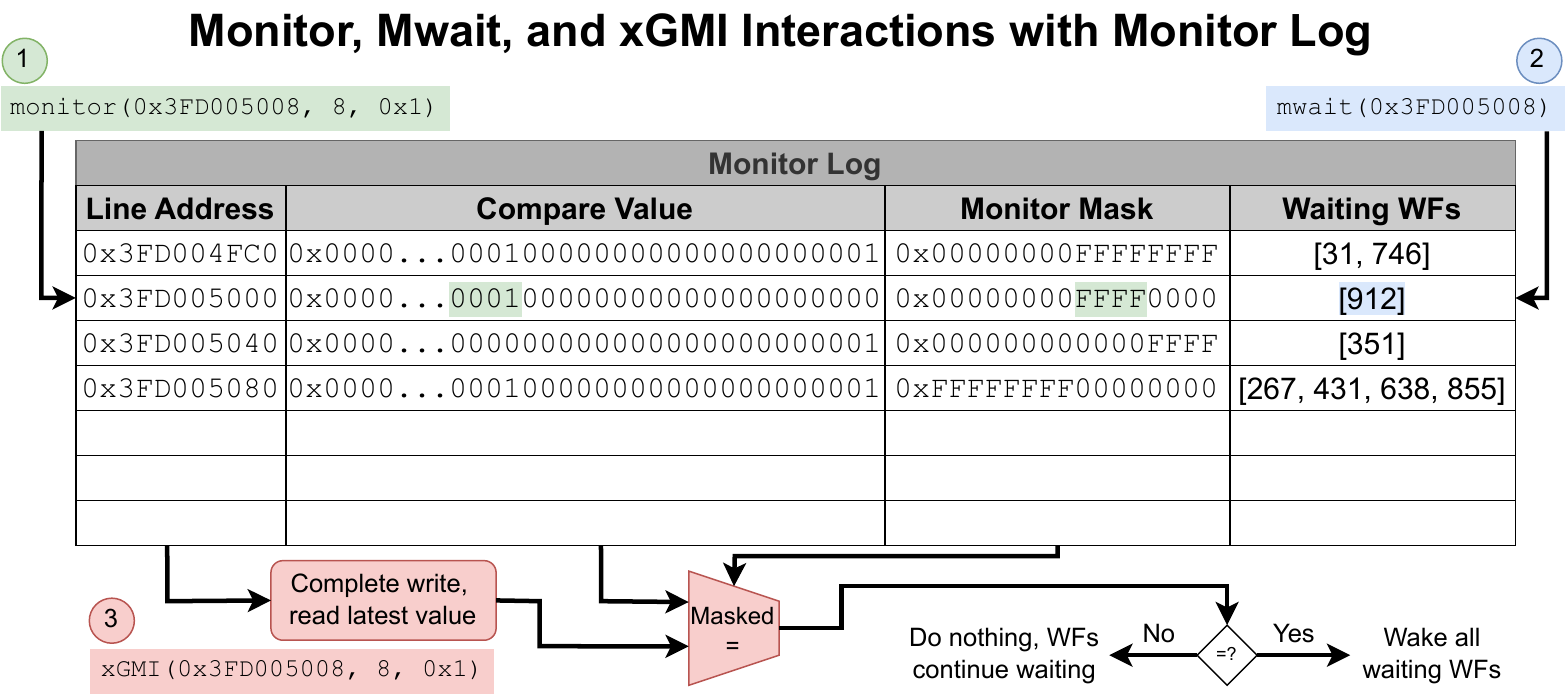}}
\caption{The Monitor Log and how the \texttt{monitor()}, \texttt{mwait()}, and emulated xGMI write events interact with this structure to trigger thread wakeups.}
\Description{The Monitor Log and how the \texttt{monitor()}, \texttt{mwait()}, and emulated xGMI write events interact with this structure to trigger thread wakeups.}
\label{fig:monitor_log}
\end{figure}

\begin{figure*}[!htb]
\centerline{\includegraphics[width=\linewidth]{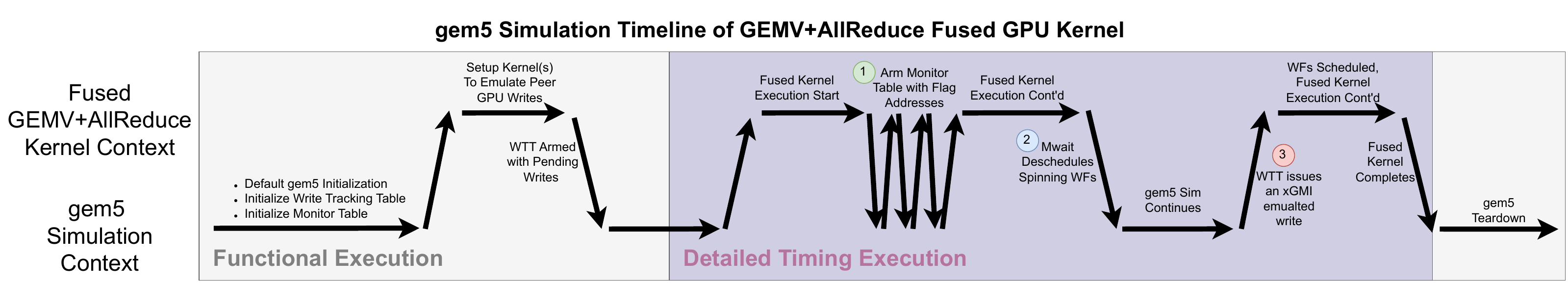}}
\caption{A timeline of the simulated execution of the fused GEMV+AllReduce kernel, annotated to show the instrumentation points used to evaluate SyncMon-inspired synchronization features within the proposed simulation framework.}
\Description{A timeline of the simulated execution of the fused GEMV+AllReduce kernel, annotated to show the instrumentation points used to evaluate SyncMon-inspired synchronization features within the proposed simulation framework.}
\label{fig:timeline}
\end{figure*}

We extend our simulator to incorporate the core concepts of SyncMon, namely the Monitor Log, which we implement as a simulator-side data structure rather than allocating it within the device’s memory space. This design decision reflects the purpose of this case study: not to propose new architectural mechanisms or evaluate hardware feasibility, but to demonstrate the flexibility and extensibility of our simulation infrastructure. The gem5 framework serves primarily as a research exploration tool, and in that spirit, our goal is to show that with the proposed enhancements to gem5, complex microarchitectural concepts such as SyncMon can be replicated, explored, and scaled across multi-GPU systems with minimal effort. By modeling the Monitor Log within the simulator state, we can easily control and observe its parameters, validate synchronization behavior, and characterize the effects of scaling out to larger GPU networks. The intent is not to reproduce a facsimile of the full SyncMon microarchitecture at hardware fidelity, but rather to illustrate that key ideas in GPU interconnect and synchronization research can be rapidly prototyped within our framework.

Through this implementation, GPU threads can register monitored memory locations and suspend execution until a peer GPU write is detected at that address. Once a write is observed, modeled as an xGMI write as described in Section \ref{SetupKernelSection}, the simulator triggers the corresponding monitor entry, resuming the waiting wavefronts. This mechanism reduces unnecessary read traffic and more accurately captures the performance characteristics of spin-yield synchronization in \cite{kishore_SC24}.

To simulate this behavior, we introduce two new GPU pseudo operations, \texttt{monitor()} and \texttt{mwait()}, which extend the gem5 instruction interface in a manner similar to the \texttt{register\_write()} function described earlier in Section \ref{SetupKernelSection}. Together, these pseudo ops emulate the semantics of the x86 \texttt{monitor} and \texttt{mwait} instructions, allowing a GPU thread to register interest in a specific memory location and suspend execution until an update is detected. 

Our \texttt{monitor()} GPU pseudo op takes three arguments: \texttt{addr}, \texttt{numBytes}, and \texttt{wakeValue}. The \texttt{addr} argument specifies the starting address of the memory region to be monitored, while \texttt{numBytes} indicates the size of the flag or synchronization variable. This size flexibility accommodates padded flags used to prevent false sharing. The \texttt{wakeValue} argument defines the expected value of the monitored flag that will trigger a wake-up event, allowing the same mechanism to represent a wide range of synchronization primitives such as mutexes, semaphores, and barriers. This design extends beyond the x86 monitor instruction, which simply observes writes within a fixed 8-byte range from a register-specified address. By allowing for explicit specification of monitored size and wake conditions, the GPU \texttt{monitor()} pseudo op provides a richer and more controllable interface for exploring synchronization mechanisms.

The \texttt{mwait()} pseudo op complements \texttt{monitor()} by suspending execution of a calling wavefront until one of its monitored memory addresses meets a wakeup condition. It takes a single argument, \texttt{addr}, the address being waited on, and records the calling wavefront’s ID in the corresponding Monitor Log entry. Once issued, the wavefront is marked to be descheduled by the GPU scheduler, freeing compute resources for other active wavefronts. During simulation, each memory write that completes at the cache directory is compared against the entries in the Monitor Log. If a matching address is found and its associated \texttt{wakeValue} condition is satisfied, all wavefronts waiting on that entry are marked as ready and rescheduled by the command processor in the subsequent dispatch cycle.

This design allows the simulator to model spin-yield behavior efficiently and deterministically, capturing the intended functionality of the SyncMon hardware described in \cite{ifp_sync_mon}. More importantly, it demonstrates that complex synchronization schemes can be prototyped and evaluated in gem5 with minimal instrumentation and simulator overhead, providing researchers with a flexible environment to explore new ideas.

The \texttt{monitor()} operation \textcircled{\scriptsize 1} initializes an entry in the simulator-side Monitor Log, shown in Figure \ref{fig:monitor_log}, with the address of the monitored variable, the size of the monitored region (stored in the monitor mask), and the expected value that will trigger a wake-up. When the instrumented application issues an \texttt{mwait()} operation for the same address \textcircled{\scriptsize 2}, the wavefront ID of the calling thread is recorded in the corresponding Monitor Log entry, and the wavefront is descheduled by the GPU scheduler. Multiple wavefronts may register to the same table entry when different synchronization flags share a cache line, as the instrumentation does not impose any ordering or exclusivity among flags.

Execution resumes only when an emulated xGMI write to the same cache line occurs \textcircled{\scriptsize 3}. Upon such a write, the simulator performs a masked comparison between the written data and the stored wake-up value. If a match is detected, all waiting wavefronts are marked as schedulable. Figure \ref{fig:timeline} presents a timeline of the instrumented fused GEMV+AllReduce kernel. The simulation begins in fast functional mode to initialize the gem5 structures and any setup kernels, and then transitions to detailed timing mode to collect fine-grained synchronization and communication statistics.

Eidola leverages common profiling timing data from the standard training lifecycle of distributed machine learning workloads. Lightweight timing profiles are collected from real executions to capture inter-GPU communication behavior. For the fused GEMV+AllReduce kernel and the other kernels from \cite{kishore_SC24}, peer-to-peer GPU write timestamps fully encapsulate the network traffic. These timestamps are registered into the simulator before kernel execution using a GPU pseudo-op that schedules future emulated inter-GPU writes. During gem5’s execution-driven simulation of a single target GPU, executing in detailed timing mode, all other GPUs are abstracted as lightweight eidolons (eGPUs) that replay recorded communication events.

The framework allows researchers to explore different wake-up granularities: at a coarse level, all waiting wavefronts can be woken following Mesa-style synchronization semantics (mwait must be called within the while loop from lines \texttt{15}–\texttt{16} in Figure \ref{fig:gemv_allreduce}, while finer-grained tracking of individual flags can be implemented to emulate Hoare-style semantics. Again, this case study is not intended to reproduce SyncMon \cite{ifp_sync_mon} at hardware fidelity, but rather to demonstrate how our enhanced simulation framework enables rapid, scalable exploration of advanced GPU synchronization mechanisms in multi-GPU systems.

\subsection{Enabling Architectural Research}\label{sec:enabling_arch_research}

We evaluate this extension in our multi-GPU simulation environment under the same test configuration described in \ref{sec:results}. As shown in Figure~\ref{fig:syncmon_results}, once SyncMon support is enabled, the number of memory reads issued by the simulated GPU no longer scales linearly with the modeled wait time. This contrasts sharply with the baseline spin-wait behavior shown in Figure~\ref{fig:configurability}, where increased wait times led to a proportional increase in polling reads due to continuous flag checking.

With SyncMon-enabled execution, the number of reads associated with polling on flag variables remains effectively constant across all configurations, ranging between 728 and 788 reads. This is a significant departure from the baseline, where polling traffic grows with the duration of the wait. The bounded number of flag-related reads reflects the transition from an active spin-wait to a spin-yield synchronization model, where waiting wavefronts are descheduled and only resume execution upon receiving a notification event.

Non-flag-related memory reads remain largely unchanged, staying consistent with the values observed in Figure~\ref{fig:configurability} at approximately 66K reads. This indicates that the introduction of SyncMon does not perturb the underlying computational behavior or memory access patterns of the kernel. Instead, it selectively reduces only the unnecessary memory traffic generated by synchronization, isolating the benefit of the proposed mechanism. These results validate the expected behavior described in the original SyncMon proposal and demonstrate that Eidola can accurately capture the performance implications of advanced synchronization mechanisms.

\begin{boxE}
\textbf{Key Takeaway:} Eidola enables detailed evaluation of communication and synchronization mechanisms.
\end{boxE}

\subsection{Simulation Scaling with Input Size}\label{sec:sim_time_scaling_input}

To evaluate how Eidola behaves under increasing computational workload, we study the relationship between application input size and overall simulation time. In particular, we vary the input matrix dimension \texttt{M}, which corresponds to the number of output rows computed by the fused GEMV+AllReduce kernel. Increasing \texttt{M} proportionally increases the amount of arithmetic work performed by each GPU, as more multiply-accumulate operations must be completed per kernel invocation. This experiment allows us to verify that Eidola preserves expected scaling trends as application complexity grows.

\begin{figure}[!htbp]
\centerline{\includegraphics[width=\linewidth]{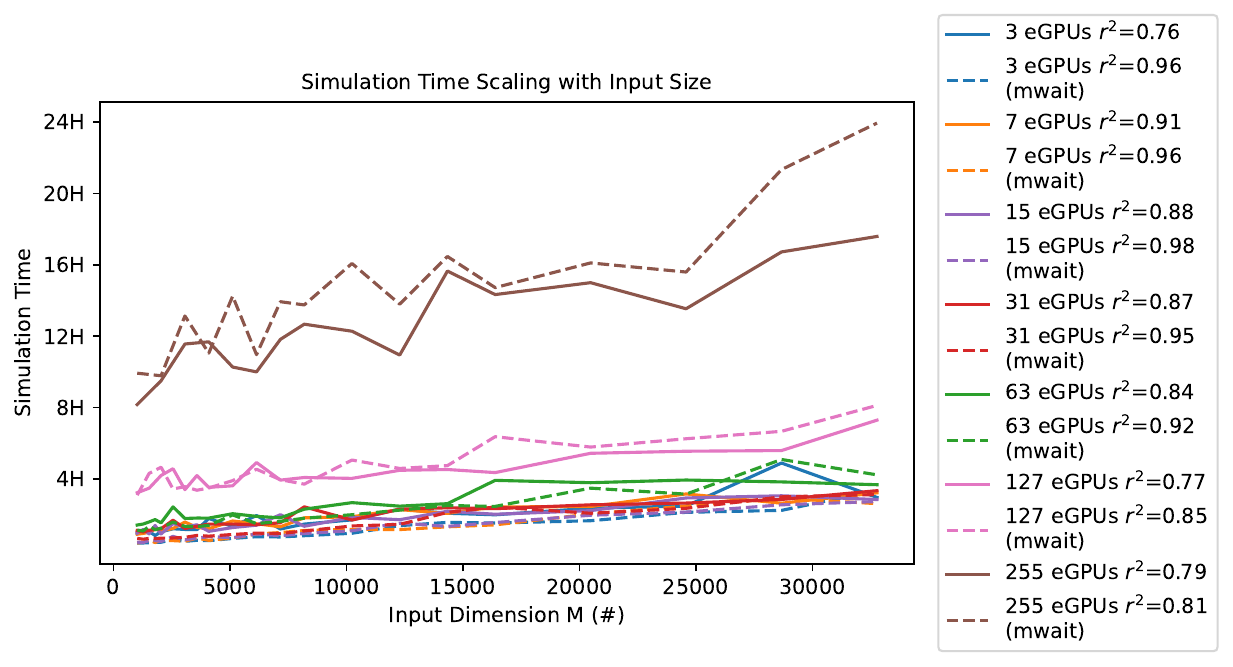}}
\caption{Plots showing gem5 simulation time of the GEMV+AllReduce kernel with varying input matrix dimension \texttt{M}. Vertical axis shows wall clock time of gem5 simulations. Horizontal axis represents number of rows simulated in input matrix. Solid lines show results from default application with emulated xGMI writes. Dashed lines show results from mwait-instrumented application.}
\Description{Plots showing gem5 simulation time of the GEMV+AllReduce kernel with varying input matrix dimension \texttt{M}. Vertical axis shows wall clock time of gem5 simulations. Horizontal axis represents number of rows simulated in input matrix. Solid lines show results from default application with emulated xGMI writes. Dashed lines show results from mwait-instrumented application.}
\label{fig:input_size_scaling}
\end{figure}

Figure~\ref{fig:input_size_scaling} compares total gem5 simulation time as the input dimension \texttt{M} is varied. Across all configurations, simulation time increases approximately linearly with \texttt{M}, as indicated by $r^2$ values ranging from 0.76 to 0.98 for the fitted trendlines. This behavior is consistent with expectations: as \texttt{M} grows, each GPU executes a larger number of compute operations, leading to a proportional increase in detailed timing simulation.

Importantly, both the baseline configuration (solid lines) and the mwait-instrumented configuration (dashed lines) exhibit similar scaling trends. This indicates that the additional synchronization modeling introduced by Eidola, including emulated xGMI writes and optional mwait-based behavior, does not fundamentally alter the relationship between workload size and simulation cost. Instead, the dominant factor remains the execution time of the simulated GPU pipeline. These results confirm that Eidola preserves the expected computational scaling characteristics of the underlying application while introducing only modest overhead for communication modeling. 

\begin{boxE}
\textbf{Key Takeaway:} Eidola preserves expected application-level scaling.
\end{boxE}

\subsection{Simulation Scaling with Emulated GPUs} \label{sec:sim_time_scaling_egpu}

The fused GEMV+AllReduce kernel inherently requires multiple GPUs to execute, as its computation depends on distributed data exchange and synchronization across devices. In practice, the kernel requires at least two GPUs to run and achieves its highest efficiency when executed on a power-of-two number of GPUs. As a result, directly measuring the simulation time of a single GPU execution ($t_{1GPU_M}$) is not possible for this workload. This presents a challenge when attempting to quantify the overhead introduced by Eidola relative to a baseline single-GPU detailed simulation.

To address this, we vary the number of emulated GPUs (eGPUs) from 3 to 255 and measure the total wall-clock simulation time across configurations. We then use these measurements to extrapolate the effective cost of simulating a single GPU and the incremental overhead associated with each additional eGPU. Specifically, we group trials by input dimension $M$, fit a linear model across configurations, and extrapolate to $eGPU = 1$ to estimate both $t_{1GPU_M}$ and $t_{eGPU_M}$. This approach allows us to isolate the contribution of Eidola’s communication modeling infrastructure to overall simulation time, even in the absence of a directly measurable single-GPU baseline.

Formally, we model total simulation time $t_M$ as:
\begin{equation} \label{eqn:linear_model}
t_{M} = t_{1GPU_{M}} + eGPUs * t_{eGPU_M}
\end{equation}
where $t_{1GPU_M}$ represents the estimated time to simulate a single GPU in detailed timing mode, and $t_{eGPU_M}$ captures the per-eGPU overhead associated with maintaining additional communication state and replaying inter-GPU events. Here, $t_M$ is the measured simulation time for a given configuration, and $eGPUs$ is the number of emulated GPUs.

\begin{figure}[htbp]
\centerline{\includegraphics[width=\linewidth]{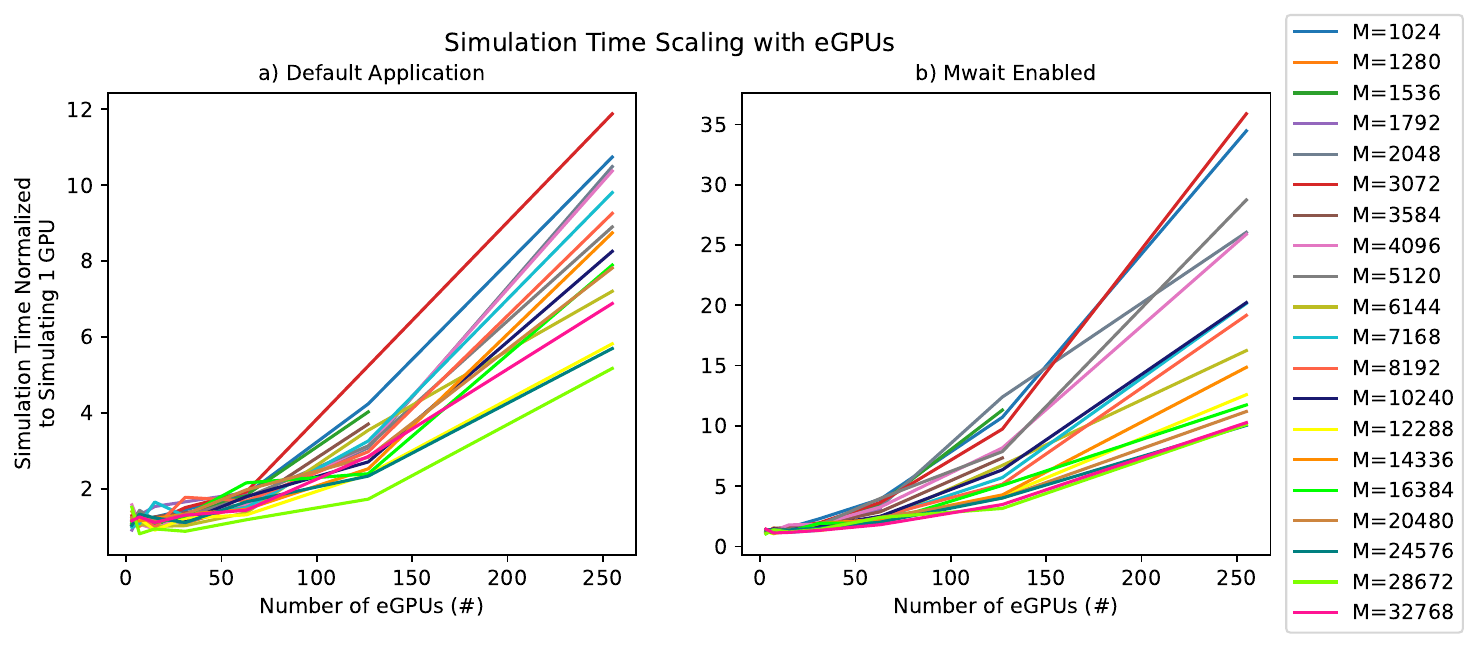}}
\caption{gem5 simulation time as number of eGPUs varies. Vertical axis represents relative simulation time normalized to estimation of simulating single GPU. Horizontal axis varies the number of emulated GPUs. Figure \ref{fig:eGPU_scaling}\textbf{a)} shows results from the application with xGMI emulated write instrumentation alone, while Figure \ref{fig:eGPU_scaling}\textbf{b)} shows results including mwait-instrumentation.}
\Description{gem5 simulation time as number of eGPUs varies. Vertical axis represents relative simulation time normalized to estimation of simulating single GPU. Horizontal axis varies the number of emulated GPUs. Figure \ref{fig:eGPU_scaling}\textbf{a)} shows results from the application with xGMI emulated write instrumentation alone, while Figure \ref{fig:eGPU_scaling}\textbf{b)} shows results including mwait-instrumentation.}
\label{fig:eGPU_scaling}
\end{figure}

Because simulation time is also a function of input size, all parameters are derived per input dimension \texttt{M}, as discussed in Section~\ref{sec:sim_time_scaling_input}. After fitting the linear model in Equation~\ref{eqn:linear_model}, we normalize the measured simulation time by the estimated $t_{1GPU_M}$ and plot the results in Figure~\ref{fig:eGPU_scaling}. This normalization enables direct comparison against an idealized baseline in which only a single GPU is simulated in detail.

Figures~\ref{fig:eGPU_scaling}\textbf{a)} and \ref{fig:eGPU_scaling}\textbf{b)} show that simulation time grows sub-linearly with the number of eGPUs across all configurations. For small numbers of eGPUs, normalized execution time approaches 1.0, indicating that the simulation cost is dominated by the detailed modeling of the primary GPU. As the number of eGPUs increases, total simulation time increases, but at a much slower rate than linear scaling. Even in the largest configuration with 255 eGPUs and mwait instrumentation enabled, normalized execution time ranges between 7.3$\times$ and 35.9$\times$, far below the 256$\times$ cost that would be expected if all GPUs were simulated in full detail.

This behavior confirms that the overhead introduced by Eidola is modest relative to the cost of detailed GPU simulation. The estimated $t_{eGPU_M}$ values indicate that the additional work required to maintain the Write Tracking Table (WTT) and replay communication events is small and scales sub-linearly with the number of eGPUs. While an event-driven implementation using gem5’s native event queues could further reduce this overhead by eliminating per-cycle polling, the current design already provides a favorable trade-off between efficiency, transparency, and ease of debugging.

Overall, these results demonstrate that Eidola enables scalable multi-GPU simulation by avoiding the need to fully model every GPU in the system, while still capturing the essential communication behavior required for architectural analysis.

\begin{boxE}
\textbf{Key Takeaway:} Eidola introduces modest simulation overheads as system size increases.
\end{boxE}

\subsection{Case Study Summary}

The SyncMon study exemplifies how our proposed methodology can be directly applied to support GPU architectural research. Across the preceding experiments, we demonstrate that Eidola enables both accurate modeling of communication-driven behavior and scalable simulation of large multi-GPU systems. In Section~\ref{sec:enabling_arch_research}, we validated Eidola’s ability to model architectural mechanisms through the SyncMon case study. By replacing spin-wait polling with a notification-based synchronization primitive, we observed that flag-related memory reads remain bounded (728–788 reads) rather than scaling with wait time, while non-flag memory traffic remains stable at approximately 66K reads. This demonstrates that Eidola can isolate and quantify the impact of synchronization mechanisms on memory behavior without perturbing the underlying computation.

In Section~\ref{sec:sim_time_scaling_input}, we showed that simulation time scales linearly with input size, confirming that Eidola preserves the expected relationship between application workload and execution cost. This ensures that increasing computational intensity does not introduce non-physical artifacts in simulation, allowing researchers to confidently study workload-dependent trends.

Finally, in Section~\ref{sec:sim_time_scaling_egpu}, we further demonstrated that simulation time grows sub-linearly with the number of emulated GPUs. By extrapolating a single-GPU baseline and isolating the per-eGPU overhead, we showed that Eidola avoids the prohibitive cost of fully simulating all devices while still capturing inter-GPU communication behavior. Even at large scales, the additional overhead introduced by communication replay and Write Tracking Table (WTT) management remains modest relative to detailed GPU pipeline simulation. This scalability is critical for studying modern distributed workloads, which routinely span tens to hundreds of GPUs.

Taken together, these results highlight Eidola’s effectiveness as a research platform. By providing cycle-level, configurable modeling of multi-GPU systems, Eidola opens the door to exploring emerging hardware features in realistic, large-scale environments. As multi-GPU systems continue to underpin large-scale model training, such capabilities will be increasingly important for guiding both academic research and industrial design.

\section{Related Work}\label{sec:related_work}

Several simulation frameworks have explored GPU and heterogeneous system modeling, but they target different aspects than our work. Multi2Sim \cite{multi2sim_2012} focuses on CPU–GPU interaction and host-driven communication, and does not model peer-to-peer GPU communication that dominates modern distributed training workloads. GPGPU-Sim \cite{gpgpu_sim} is a detailed simulator designed to study single-GPU microarchitectural behavior, but it lacks native support for multi-GPU systems and inter-GPU synchronization. 

MGPUSim \cite{mgpusim_yifan_sun_2019} provides a detailed multi-GPU simulation environment and is the closest to our work; however, it models all GPUs in a cluster at fine granularity. In contrast, Eidola abstracts the behavior of non-target GPUs while preserving detailed simulation for selected devices, enabling scalable modeling of inter-GPU communication traffic and synchronization effects in large multi-GPU systems. Eidola and MGPUSim address different points in the multi-GPU simulation design space. MGPUSim provides detailed, event-driven simulation of all GPUs, which is valuable for full-system studies but limits scalability as GPU count increases. Eidola instead abstracts non-target GPUs and applies detailed timing simulation only to a selected device, enabling scalable exploration of inter-GPU communication and synchronization effects on GPU microarchitecture. This is particularly effective for computationally balanced workloads such as those in \cite{kishore_SC24}, where examining a single GPU can provide system-level insights.

While both simulators rely on event-based mechanisms, Eidola’s key contribution is decoupling execution detail from communication fidelity. By replaying cycle-level communication events from abstracted GPUs, Eidola supports simulations with hundreds of GPUs while retaining fine-grained visibility into a target GPU. 
Finally, MGPUSim models AMD GPUs based on GCN3, whereas Eidola targets modern CDNA-based GPUs (MI100, MI200, MI300), representative of current hyperscale systems.

\section{Future Work}\label{sec:future_work}

An important direction for future work is extending validation to larger, scale-out multi-GPU systems. While the timing profiles used in this study were collected from a single node containing four GPUs, our simulator modifications are designed to support configurations with tens to hundreds of GPUs. Evaluating the methodology across more complex intra-node and inter-node topologies would provide greater insight into how communication patterns scale across distributed environments. This extended validation would further establish the simulator as a reliable platform for studying emerging architectural solutions in high-performance, distributed GPU systems.

The fused GEMV+AllReduce kernel examined in this study is a balanced workload. We selected it because it stresses fine-grained synchronization, which is central to modern transformer training. The same work [30] also presents other fused kernels, such as embedding pooling + All-to-All and GEMM + All-to-All, which can be evaluated using Eidola without modification. Importantly, Eidola imposes no constraints on workload balance. User-specified profiling enables support of arbitrary and asymmetric communication patterns, ideal for workloads exhibiting producer–consumer data patterns. It is our hope that Eidola will enable research spanning a broad diversity of GPU workloads that exploit both heterogeneity and asymmetry.

\section{Conclusion}\label{sec:conclusion}

The growing demand for large-scale machine learning has made multi-GPU systems central to modern training infrastructure, yet understanding the performance impact of inter-GPU communication remains challenging. This work extends gem5 to provide configurable, timing-accurate modeling of multi-GPU systems, Eidola, using real application traces annotated with precise timestamps to emulate peer-to-peer GPU writes.

Eidola reproduces communication-induced memory traffic with fine-grained control, enabling detailed analysis of synchronization delays. We validate this capability by demonstrating linear scaling of spin-wait traffic with write delay and by implementing key features of the SyncMon proposal, where spin-yield synchronization reduces memory reads as expected. Eidola represents an initial step toward scalable simulation of multi-GPU workloads with fine-grained synchronization. The case study in Section \ref{sec:case_study_sync_mon} demonstrates both its scalability and usefulness.

These results show that our approach supports realistic modeling of inter-GPU interactions and facilitates architectural exploration of emerging synchronization mechanisms. As distributed training continues to scale, such tools will be essential for bridging device-level design and system-level performance modeling.



\bibliographystyle{ACM-Reference-Format}
\bibliography{refs}

\end{document}